\documentstyle[multicol,prb,aps]{revtex}

\begin{document}
\draft
\title{Spin injection into a ballistic semiconductor microstructure}
\author{Vladimir Ya. Kravchenko$^{1,4}$ and Emmanuel I.~Rashba$^{2,3,}$\cite{Rashba*}}
\address{$^1$Argonne National Laboratory, 9700 S. Cass Avenue, Argonne, IL 60439\\
$^2$Department of Physics, SUNY at Buffalo, Buffalo, NY 14260\\
$^3$Department of Physics, MIT, Cambridge, Massachusetts 02139\\
$^4$Institute of Solid State Physics, Chernogolovka, Moscow District, 142432 Russia}
\date{September 9, 2002}
\maketitle
\begin{abstract} 
  A theory of spin injection across a ballistic ferromagnet-semiconductor-ferromagnet junction is developed for the Boltzmann regime. Spin injection coefficient $\gamma$ is suppressed by the Sharvin resistance of the semiconductor $r_N^*=(h/e^2)(\pi^2/S_N)$, where $S_N$ is the Fermi-surface cross-section. It competes with the diffusion resistances of the ferromagnets $r_F$, and $\gamma\sim r_F/r_N^*\ll 1$ in the absence of contact barriers. Efficient spin injection can be ensured by contact barriers. Explicit formulae for the junction resistance and the spin-valve effect are presented.           
\end{abstract}
\begin{multicols}{2}

Efficient spin injection from ferromagnetic metals into semiconductors is one of the prerequisites for developing spintronics of the hybrid metal-semiconductor devices.\cite{review} As distinct from the spin injection into paramagnetic metals,\cite{JS85} the first attempts in achieving spin injection into semiconductors failed. In the framework of the diffusion theory, this failure was explained by Schmidt {\it et al.} in terms of the ``conductivity mismatch",\cite{Sch00} the spin injection coefficient $\gamma$ being of the order of magnitude $\gamma\sim r_F/r_N$ where $r_F$ and $r_N$ are the diffusive resistances of a ferromagnet and of a normal conductor (semiconductor microstructure), respectively. Resistive spin-selective contacts had been proposed to remedy the problem,\cite {R00} and an impressing progress in the experimental work was achieved during the last year.\cite{exper} The future theoretical work performed in the framework of the diffusion approximation also substantiated this approach.\cite{SS01,FJ01,R02}

The spin transistor proposed by Datta and Das\cite{DD90} and similar devices\cite{SimDiv} rely on a ballistic rather than diffusive transport. Therefore, a lot of theoretical work was performed on a coherent ballistic transport through the contacts and the microstructure confined between them. The role of the barriers at the interfaces, the Fresnel-type relations between propagating and reflected waves originating due to the parameter mismatch, and the interference pattern in the bulk caused by the spin-orbit interaction were considered.\cite{ballistic} The critical role of a barrier at the interface was emphasized\cite{barrier} and the scattering in the bulk was discussed.\cite{scat} Spin filtering through perfectly matched interfaces was investigated.\cite{perfect}

In the theory of spin injection into semiconductors there exists a gap between the pictures of a diffusive transport and of a coherent transport across the interface. To close this gap, we consider an intermediate regime when (i) the phase coherence at the interfaces is broken and (ii) electrons can be described by the Boltzmann equation inside the microstructure. Solving this equation in a nearly ballistic regime permitted us to find explicit expressions for the basic parameters of the ferromagnet-semiconductor-ferromagnet system. The importance of the problem is emphasized by recent experimental findings of Ramsteiner {\it et al.} who investigated spin injection through a high quality MnAs/GaAs interface.\cite{Ram02} They looked for the effect of the symmetry matching of the wave functions of MnAs and GaAs and found no correlation between the spin injection and the azimuthal orientation of the surface layers. Apparently, this result reflects the limitations of the current technologies.

We show that the spin injection into a ballistic semiconductor through a diffusive interface is controlled by the {\it Sharvin resistance\cite{Sh65} of the semiconductor} $r_N^*$ that solely depends on the electron concentration and the resistance quantum $h/e^2$. In the absence of the contact resistance, the spin injection coefficient $\gamma\sim r_F/r_N^*\ll 1$. Therefore, {\it contact resistances are indispensable for supporting spin injection} from a metal into a semiconductor, similar to the diffusion regime but with a different criterion and because of somewhat different arguments.

{\it Model.} We consider a normal (N, non-ferromagnetic) conductor residing in a region  $-d/2<x<d/2$ and separated by barriers from two semi-infinite ferromagnetic (F) electrodes, $|x|>d/2$. Spin-orbit interaction is neglected, all conductors are assumed degenerate, and the contacts are both spin-selective and spin-conserving. The scattering at the contacts is diffusive in concord with the assumption of the phase breaking barriers establishing a Boltzmann regime. The Boltzmann equation is solved separately in the N- and F-regions with a proper account of nonequilibrium spins and the boundary conditions at the contacts. We also assume that both contacts and ferromagnets are identical, but the polarization of the ferromagnets may be either parallel (P) or antiparallel (AP), the P- and AP-geometries in what follows.

{\it N-region, $|x|<d/2$.} In the linear approximation in the electrical field $E(x)=-\partial_x\varphi(x)$, electrons in the N-region can be described by the distribution functions $f_\alpha(x, v_x)=f_0+(df_0/d\epsilon) \psi_\alpha(x,v_x)$ where $\alpha$ is the spin index and $\psi_\alpha(x,v_x)$ satisfies the Boltzmann equation
\[v_x\partial_x\psi_\alpha(x,v_x)-eE(x)v_x+[\psi_\alpha(x,v_x)-{\bar\psi}_\alpha(x)]/\tau_N=0,\]
where $\tau_N$ is the momentum relaxation time, and ${\bar\psi}_\alpha(x)$ is the average value of $\psi_\alpha(x,v_x)$ over the Fermi sphere
${\bar\psi}_\alpha(x)=\langle\psi_\alpha(x,v_x)\rangle/\rho_\alpha$. Here $\rho_\alpha=\rho_N/2$ are the densities of states of $\alpha$-electrons, $\rho_N$ being the total density of states. Concentrations of nonequilibrium spins are equal to  $n_\alpha(x)=-\rho_N {\bar\psi}_\alpha(x)/2$. Spin relaxation is neglected. 

It is convenient to eliminate the electrical potential $\varphi(x)$ by introducing new functions
$e\zeta_\alpha(x,v_x)=-[\psi_\alpha(x,v_x)+e\varphi(x)],$
$e\zeta_\alpha(x)=-[\bar\psi_\alpha(x)+e\varphi(x)]$
that obey the equation
\begin{equation}
\zeta_\alpha^{\prime}(x,v_x)+[\zeta_\alpha(x,v_x)-\zeta_\alpha(x)]/{\tau v_x}=0.
\label{eq3}
\end{equation}
When the function $\zeta_\alpha(x)$ changes smoothly, it has a meaning of the electrochemical potential of the electrons with the spin $\alpha$. As in  similar problems with several groups of carriers,\cite{KR69,K02} one should first consider $\zeta_\alpha(x)$ as known functions, find $\zeta_\alpha(x,v_x)$, and then impose the self-consistency condition
$\zeta_\alpha(x)=\langle\zeta_\alpha(x,v_x)\rangle/\rho_\alpha$.

Similarly to the problems of the transport of radiation\cite{MF} and electrons\cite{FS} in restricted areas, Eq.~(\ref{eq3}) should be solved separately for $v_x>0$ and $v_x<0$. However, because of the symmetry of the problem, the solutions obey the relations
\begin{equation}
\zeta_\alpha(x,v_x)=-\zeta_{\bar\alpha}(-x,-v_x),~~
\zeta_\alpha(x)=-\zeta_{\bar\alpha}(-x),
\label{eq5}
\end{equation}
where $\bar\alpha=\alpha$ in the P-geometry, and $\bar\alpha=-\alpha$ in the AP-geometry. Therefore, it is enough to solve Eq.~(\ref{eq3}) for $v_x<0$. For diffusive scattering at the contacts
\begin{eqnarray}
\zeta_\alpha(x,v_x&<&0)=
\zeta_\alpha^<(N)\exp\left(-{{d/2-x}\over{\tau_N |v_x|}}\right)\nonumber\\
&-&{1\over{\tau_N |v_x|}}\int_{d/2}^xdu~\zeta_\alpha(u)\exp\left(-{{u-x}\over{\tau_N |v_x|}}\right).
\label{eq6}
\end{eqnarray}
Here $\zeta_\alpha^<(N)$ are the integration constants that are equal to the boundary values of $\zeta_\alpha(x,v_x<0)$ at the right boundary of the N-region, $x=d/2$.

For the P-geometry, an integral equation for $\zeta_\alpha(x)$ follows from Eqs.~(\ref{eq5}) and (\ref{eq6}): 
\begin{eqnarray}
\zeta_\alpha(x)&=&{\case 1/2}\zeta^<_\alpha(N)\{E_2[(d/2-x)/l_N]-E_2[(d/2+x)/l_N]\}\nonumber\\
&+&{1\over{2l_N}}\int_{-d/2}^{d/2}du~
\zeta_\alpha(u)E_1\left({{|x-u|}\over{l_N}}\right),
\label{eq7}
\end{eqnarray}
where functions $E_n(\xi)=\int_1^\infty dt~e^{-t\xi}/t^n$,\cite{MF} $l_N=\tau_Nv_N$ is the mean free path, and $v_N$ is the Fermi velocity. Function $E_1(\xi)$ is easily related to the integral exponent function, $E_1(\xi)=-{\rm Ei}(-\xi)$, $\xi>0$. Eq.~(\ref{eq7}) can be solved in the ballistic regime, $d\ll l_N$, in powers of the small parameter $\lambda=(d/l_N)\ln(l_N/d)\ll 1$. Under these conditions the integral term is small compared to the left hand side and can be disregarded. The function $E_2(\xi)=\exp(-\xi)+\xi{\rm Ei}(-\xi)$, and the expansion ${\rm Ei}(-\xi)\approx\ln\xi+{\bf C}$, where ${\bf C}\approx 0.577$ is the Euler constant, can be employed.\cite{GR} Finally, $\zeta_\alpha(N)\equiv\zeta_\alpha(x=d/2)$ equals
 $\zeta_\alpha(N)=\zeta^<_\alpha(N)(d/2l_N)[1-{\bf C}+\ln(l_N/d)].$
Therefore, asymmetric parts of the distribution functions [that are scaled by $\zeta^<_\alpha(N)$] are large compared to their symmetric parts, $\zeta_\alpha(N)$,  related to the concentrations, $n_\alpha(x)$.

In a similar way, equations for spin polarized currents $j_\alpha(x)$ can be derived. Because $j_\alpha(x)$ are conserved inside the N-region, it is enough to calculate $j_\alpha(N)\equiv j_\alpha(d/2)$
\begin{eqnarray}
j_\alpha(&N&)={\case 1/4}e^2\rho_Nv_N\{\zeta_\alpha^<(N)[E_3(0)+E_3(d/l_N)]\nonumber\\
&+&{1\over l_N}\int_{-d/2}^{d/2}du~\zeta_\alpha(u)E_2\left({{|x-u|}\over{l_N}}\right){\rm sign}(u-x)\}.
\label{eq9}
\end{eqnarray}
Omitting the integral term that is small, and calculating the first term in the leading order in $\lambda$, we find $j_\alpha(N)=e^2\rho_Nv_N\zeta_\alpha^<(N)/4$. Expressing the currents $j_\alpha(N)$ through the total current $J$ and the spin injection coefficient $\gamma_N=[j_\uparrow(N)-j_\downarrow(N)]/J$, we come to the final equation
 \begin{equation}
\zeta_\uparrow^<(N)-\zeta_\downarrow^<(N)=2r_N^*\gamma_NJ,~~
r_N^*=2/e^2\rho_Nv_N. 
\label{eq10}
\end{equation}
We show below that $r_N^*$ plays a role of the resistance of each end of the N-region (per unit area). It depends neither on $d$ or $l_N$ but solely on the electron concentration\cite{ln}
  \begin{equation}
r_N^*=(h/e^2)(\pi^2/S_N)=(h/e^2)(1/2{\cal N}), 
\label{eq11}
\end{equation}
where $h/e^2$ is the resistance quantum. Here $S_N=\pi k_N^2$ is the Fermi surface cross-section, and ${\cal N}=k_F^2/2\pi$ is the number of channels per unit cross-section area, including the spin degeneracy factor. The effective resistance of a narrow diffusive region equals $d/\sigma_N$, cf. Eq. (37) of Ref.~\onlinecite{R02}; it matches $r_N^*$ at $d\sim l$. Because $r_N^*$ depends on the carrier concentration only, develops due to the electron exchange with the diffusive regions, and has the appropriate analytical form, we identify it as the Sharvin resistance of the normal conductor.\cite{Sh65} The relevance of Sharvin resistance to the perpendicular transport in layered magnetic structures was recognized by Bauer.\cite{B92}

 {\it Right F-region, $x>d/2$}. The problem of the spin injection into a semi-infinite ferromagnet can be solved when the electron mean free paths $l_\alpha$ for both spins are small compared to the spin diffusion length $L_F$, $l_\alpha\ll L_F$. Then in the narrow layer near the contact, $w=x-d/2\ll L_F$, spin relaxation can be neglected, and the selfconsistency equation for the electrochemical potentials $\zeta_\alpha(w)$ has the form similar to Eq.~(\ref{eq7}) 
\begin{eqnarray}
\zeta_\alpha(w)&-&{\case 1/2}\zeta_\alpha^>(F)~E_2(w/l_\alpha)\nonumber\\
&=&{1\over 2l_\alpha}\int_0^\infty du~\zeta_\alpha(u)~ E_1(|u-w|/l_\alpha).
\label{eq12}
\end{eqnarray}
Here $\zeta^>_\alpha(F)$ are integration constants similar to $\zeta^<_\alpha(N)$ but for the right-moving electrons. Eliminating the second term in the left hand side of Eq.~(\ref{eq12}) by the shift
 $\zeta_\alpha(x)=\eta_\alpha(x)+\zeta^>_\alpha(F)$ 
 results in a Milne equation\cite{MF}
 \begin{equation}
\eta_\alpha(w)={1\over{2l_\alpha}}\int_0^\infty du~E_1(|w-u|/l_\alpha)\eta_\alpha(u). 
\label{eq13}
\end{equation}

For $l_\alpha \ll w$, diffusion equations can be used instead of the Milne equations, and the spin relaxation can be easily taken into account. In the region $l_\alpha\ll w \ll L_F$, both Eqs.~(\ref{eq13}) and the diffusion equations hold. Therefore, they should match smoothly. For $w\gg l_\alpha$, the asymptotic form of the solution of Eq.~(\ref{eq13}) is
 \begin{equation}
\zeta_\alpha(x)\approx\sqrt{3}\eta_\alpha(0)(w/l_\alpha+q_\infty)+\zeta^>_\alpha(F), ~q_\infty\approx 0.71.
\label{eq14}
\end{equation}
The solutions of the standard diffusion equations are\cite{R02}
\begin{eqnarray}
\zeta_\uparrow(w)&=&(\sigma_\downarrow/\sigma_F)\zeta_F\exp(-w/L_F)+Jw/\sigma_F+z_R,\nonumber\\
\zeta_\downarrow(w)&=&-(\sigma_\uparrow/\sigma_F)\zeta_F\exp(-w/L_F)+Jw/\sigma_F+z_R,
\label{eq15}
\end{eqnarray}
where $\zeta_F$ and $z_R$ are integration constants, $\sigma_\alpha=e^2v_\alpha\rho_\alpha l_\alpha/3$ are the conductivities for both spins, $v_\alpha$ and $\rho_\alpha$ are their Fermi velocities and densities of states, and $\sigma_F=\sigma_\uparrow+\sigma_\downarrow$. Matching the expansion of Eqs.~(\ref{eq15}) for small $w$, $w\ll L_F$, with Eq.~(\ref{eq14}) results in
 \begin{equation}
\zeta_F=\sqrt{3}q_\infty[\eta_\uparrow(0)-\eta_\downarrow(0)]+[\zeta_\uparrow^>(F)-\zeta^>_\downarrow(F)].
\label{eq16}
\end{equation}
and
 \begin{eqnarray}
\sqrt{3}\eta_\uparrow(0)/l_\uparrow&=&J/\sigma_F -(\sigma_\downarrow/\sigma_F)(\zeta_F/L_F),\nonumber\\
\sqrt{3}\eta_\downarrow(0)/l_\downarrow&=&J/\sigma_F +(\sigma_\uparrow/\sigma_F)(\zeta_F/L_F).
\label{eq17}
\end{eqnarray}
It follows from Eq.~(\ref{eq17}) that $\eta_\alpha(0)\sim (l_\alpha/L_F)\zeta_F\ll \zeta_F $, hence, Eq.~(\ref{eq16}) reduces to $\zeta_F=\zeta_\uparrow^>(F)-\zeta^>_\downarrow(F)$.

Spin polarized currents $j_\alpha(F)$ at the boundary of the F-region can be found either as $j_\alpha(F)=e^2v_\alpha\rho_\alpha\eta_\alpha(0)/\sqrt{3}$ or as $j_\alpha(F)=\sigma_\alpha\partial_w\zeta_\alpha(w\rightarrow 0)$. Through them the spin injection coefficient $\gamma_F$ at the left boundary of the ferromagnet can be found. In the lower order in $l_\alpha/L_F\ll 1$, the final result is
 \begin{equation}
\gamma_F=\Delta\sigma/\sigma_F-[\zeta_\uparrow^>(F)-\zeta_\downarrow^>(F)]/2r_FJ,
\label{eq18}
\end{equation}
where $\Delta\sigma=\sigma_\uparrow-\sigma_\downarrow$, and $r_F=\sigma_FL_F/4\sigma_\uparrow\sigma_\downarrow$ is the diffusive resistance of the F-region.

{\it Right contact, $x=d/2$.} A tunnel or Schottky barrier separating the N- and F-regions can be described by the transparency coefficients, $t_\alpha^{NF}$ and $t_\alpha^{FN}$, for the electrons reaching the contact from its N and F sides, respectively. They are related by the detailed balance condition ${\case 1/2}t_\alpha^{NF}v_N\rho_N=t_\alpha^{FN}v_\alpha\rho_\alpha$. Spin polarized currents flowing through a spin-conserving barrier are
 \begin{equation}
j_\alpha=-e^2[t^{NF}_\alpha \langle\zeta_\alpha(N,v_x)v_x\rangle_+
+t^{FN}_\alpha \langle\zeta_\alpha(F,v_x)v_x\rangle_-]
\label{eq19}
\end{equation}
where the symbols $\langle ...\rangle_{\pm}$ indicate that the averaging should be performed only over the right or the left hemisphere, and $\zeta_\alpha(N,v_x)$ and $\zeta_\alpha(F,v_x)$ are the functions $\zeta(x, v_x)$ for the in-coming electrons at the left and the right sides of the contact, respectively. The average values appearing in Eq.~(\ref{eq19}) can be found from the equations for the currents on both sides of the junction
 \begin{eqnarray}
j_\alpha(N)&=&-e^2[\langle\zeta_\alpha(N,v_x)v_x\rangle_+
+\zeta_\alpha^<(N)\langle v_x\rangle_-],\nonumber\\
j_\alpha(F)&=&-e^2[\langle\zeta_\alpha(F,v_x)v_x\rangle_-
+\zeta_\alpha^>(F)\langle v_x\rangle_+].
\label{eq20}
\end{eqnarray}
Because of the spin conservation, $j_\alpha(N)=j_\alpha(F)=j_\alpha$, and Eq.~(\ref{eq19}) can be transformed to the form
 \begin{equation}
j_\alpha=[\zeta_\alpha^>(F)-\zeta_\alpha^<(N)]/r_\alpha,
\label{eq21}
\end{equation}
where the effective contact resistances $r_\alpha$ are
 \begin{equation}
r_\alpha=4r_N^*(1-t_\alpha^{NF}-t_\alpha^{FN})/t_\alpha^{NF},
\label{eq22}
\end{equation}
and $r_N^*$ is defined by Eq.~(\ref{eq11}).

If the contact behaves like ``a two-side black body", i.e., if it absorbs all in-coming electrons and ensures an equilibrium emission of them into both sides, then $t_\alpha^{FN}+t_\alpha^{NF}=1$. This fact follows from arguments similar to those invoked in proving the Kirchhoff theorem in the theory of radiation. Reflection from the barrier and nonequilibrium inside it result in $t_\alpha^{FN}+t_\alpha^{NF}<1$. Hence, $r_\alpha$ are  positive and can be considered as contact resistances. 

Equation for $\gamma$ following from Eq.~(\ref{eq21}) is
 \begin{eqnarray}
\gamma_c&=&\Delta r_c/{r_c}\nonumber\\
&+& \{[\zeta_\uparrow^>(F)-\zeta_\downarrow^>(F)]
-[\zeta_\uparrow^<(N)-\zeta_\downarrow^<(N)]\}/2r_cJ,
\label{eq23}
\end{eqnarray}
where $r_c=(r_\uparrow+r_\downarrow)/4$, $\Delta r_c=(r_\downarrow-r_\uparrow)/4$. The denominator 4 ensures direct connection to the diffusive theory.

{\it Spin injection coefficient $\gamma$.} Applying the $\gamma$-technique,\cite{R02} we write $\gamma_N=\gamma_F=\gamma_c=\gamma$, and Eqs.~(\ref{eq10}), (\ref{eq18}) and (\ref{eq23}) result in an equation for $\gamma$. Its solution is
 \begin{equation}
\gamma=\left[\Delta r_c+r_F(\Delta\sigma/\sigma_F)\right]/r_{FN}^*,
\label{eq24}
\end{equation}
where $r_{FN}^*=r_F+r_c+r_N^*$. It resembles the result of the diffusion theory for a F-N-F-junction: the resistance $r_{FN}^*$ in the denominator and spin-selectivities of the bulk and the contact in the numerator; cf. Eqs. (36) and (37) of Ref.\onlinecite{R02}. For a narrow junction, $d\rightarrow 0$, $r_N^*$ is standing for $d/2\sigma_N$. As the total resistance of the junction includes $2r_F$ and $2r_c$, the resistance $r_N^*$ acquires a meaning of the Sharvin resistance of each end of the ballistic region, the left and the right. It equals $\pi\hbar/e^2$ per spin channel.\cite{FSR} 

For the spin injection from a ferromagnetic metal like Co into a semiconductor microstructure, $r_N^*\agt 10^3r_F$ and with $r_c\approx\Delta r_c\approx 0$ spin injection is strongly suppressed, $\gamma\sim r_F/r_N^*\ll 1$. In the diffusive regime, a similar effect is attributed to the conductivity mismatch\cite{Sch00} because of the large diffusive resistance $r_N=L_N/\sigma_N$, $L_N$ being the spin diffusion length in the N conductor. Eq.~(\ref{eq11}) for $r_N^*$ does not include $\sigma_N$ and $L_N$, and the large value of $r_N^*$ comes solely from the {\it low electron concentration}, $n_N^{(0)}\propto k_N^3$. However, because the resistances $r_\alpha$ are scaled by $r_N^*$ and enhanced by small $t^{NF}_\alpha$, Eq.~(\ref{eq22}), the spin injection coefficient $\gamma\approx \Delta r_c/r_c$ may be large enough and is controlled by the contact rather than the bulk.\cite{SpRel} We conclude that {\it $\gamma$ is suppressed even in the ballistic regime and contact resistances are needed to enhance it.}

A similar technique can be applied to a contact embedded between semi-infinite F- and N- regions or to any kind of a ballistic spin filter.\cite{filter} It allows one to calculate $\gamma$ and to relate the contact conductivities $\Sigma_\alpha$ of the diffusion theory\cite{R00,R02} to the contact resistances $r_\alpha$ of Eq.~(\ref{eq22}). A straightforward calculation results in $\Sigma_\alpha=1/r_\alpha$. This equation establishes a connection between the parameters of the kinetic and diffusion theories.\cite{eff} 

{\it F-N-F-junction resistance.} With the electrochemical potentials $\zeta_\alpha(w)$ found above, one can calculate the integration constant $z_R$ of Eq.~(\ref{eq15}). The two-contact resistance of the junction, $R=2z_R/J$, equals:
\begin{eqnarray}
R=-2\gamma^2r_{FN}^*+2(r_c+r_N^*)+2r_F(\Delta\sigma/\sigma_F)^2.
\label{eq25}
\end{eqnarray}
Eq.~(\ref{eq25}) differers from the $d\rightarrow 0$ limit of the diffusive theory\cite{R02} by the substitution $r_c\rightarrow r_c+r_N^*$.

 Similar to the diffusive regime, $R$ can be split into the equilibrium and nonequilibrium parts,\cite{R00,R02} $R=R_{\rm eq}+R_{\rm n-eq}$, the latter part turns into zero when $r_F\rightarrow 0$ and comes from the nonequilibrium spins in the F-regions:
\begin{eqnarray}
R_{\rm eq}&=&2(r_N^*+r_\uparrow/2)(r_N^*+r_\downarrow/2)/(r_c+r_N^*),\nonumber\\
R_{\rm n-eq}&=&2r_F[\Delta r_c-(r_c+r_N^*)\Delta\sigma/\sigma_F]^2/r^*_{FN}(r_c+r^*_N).
\label{eq26}
\end{eqnarray}
In the limit $r_N^*=0$ one recovers the diffusive resistances of Ref.~\onlinecite{R02} found for $d\rightarrow 0$, cf. Eq. (45) and Appendix.

From Eqs.~(\ref{eq24}) and (\ref{eq26}), a prescription follows: {\it parameters of the ballistic theory can be found from the $d=0$ limit of the diffusive theory by plugging} $r_c\rightarrow (r_c+r_N^*)$.

{\it AP-geometry and the spin valve effect}. In this geometry $\gamma_{AP} =0$ because of the symmetry arguments. Calculating the resistance shows that $R_{AP}$ differs from $R$ of Eq.~(\ref{eq25}) only by the absence of the first term. Therefore, the spin valve effect equals $\Delta R=2\gamma^2r_{FN}^*$, where $\gamma$ is determined by Eq.~(\ref{eq24}).

In conclusion, it is the Sharvin resistance of the semiconductor microstructure that controls spin injection across a ballistic F-N-F-junction in the Boltzmann regime. This resistance is larger than the effective resistances of ferromagnetic leads, and resistive spin-selective contacts are needed to ensure efficient spin injection.

We are grateful to Dr. I. \v{Z}uti\'{c} for useful comments. V.Ya.K. acknowledges the support from U.S. DOE Office of Science under Contract No. W31-109-ENG-38, and E.I.R. the support from DARPA/SPINS by the Office of Naval Research Grant N000140010819.

\end{multicols} 

\begin{references}
\bibitem{FSR} Sharvin resistances of ferromagnets $r_\alpha^*\sim(h/e^2)/k_\alpha^2$, where $k_\alpha$ are Fermi momenta, are small as compared to $r_F$ in the parameter ${\it l}_\alpha/L_F\ll 1$ and, therefore, were neglected.
\end{references}
\end{document}